\documentclass[twocolumn,prd,aps,floatfix]{revtex4}

\usepackage{amsmath}
\usepackage{latexsym}
\usepackage[psamsfonts]{amssymb}
\usepackage{graphicx}
\usepackage{longtable}
\usepackage{epstopdf}
\usepackage{epsfig}
\usepackage{bm}
\usepackage{color}

\newcommand{\be}{\begin{equation}}
\newcommand{\ee}{\end{equation}}
\newcommand{\bea}{\begin{eqnarray}}
\newcommand{\eea}{\end{eqnarray}}
\newcommand{\bi}{\begin{itemize}}
\newcommand{\ei}{\end{itemize}}

\newcommand{\Pslash}{p \kern -2mm /}
\newcommand{\Btob}{B_1\rightarrow B_2}
\newcommand{\btoB}{B_2\rightarrow B_1}

\newcommand{\XtoS}{\Xi\rightarrow \Sigma}

\newcommand{\StoN}{\Sigma\rightarrow N}

\begin{document}

\title{Hyperon vector form factor from 2+1 flavor lattice QCD}
\author{Shoichi Sasaki} 
\email[E-mail: ]{ssasaki@nucl.phys.tohoku.ac.jp}

\affiliation{Department of Physics, Tohoku University, Sendai 980-8578, Japan}

\date{\today}
\begin{abstract}
We present the first result for the hyperon vector form factor $f_1$ 
for $\Xi^0\rightarrow \Sigma^+ l\bar{\nu}$ and $\Sigma^-\rightarrow n l\bar{\nu}$ semileptonic decays 
from fully dynamical lattice QCD. The calculations are carried out with 
gauge configurations generated by the RBC and UKQCD collaborations with (2+1)-flavors 
of dynamical domain-wall fermions and the Iwasaki gauge action
at $\beta=2.13$, corresponding to a cutoff $a^{-1}=1.73$ GeV. 
Our results, which are calculated at the lighter three sea quark masses (the lightest pion 
mass down to approximately 330 MeV), show that a sign of the second-order correction of SU(3) 
breaking on the hyperon vector coupling $f_1(0)$ is negative. 
The tendency of the SU(3) breaking correction observed in this work disagrees with predictions of 
both the latest baryon chiral perturbation theory result and large $N_c$ analysis. 
\end{abstract}


\maketitle

\newpage

 
\section{Introduction}
Greater knowledge of the vector form factor $f_1$ in $\Delta S=1$ semileptonic hyperon decays
paves the way for an alternative determination of the 
element $V_{us}$ of the Cabibbo-Kobayashi-Maskawa (CKM) matrix in addition to
kaon semileptonic ($K_{l3}$) decays, leptonic decays of kaons and pions, and 
hadronic decays of $\tau$ leptons~\footnote{Recently, first principles calculations of the $K_{l3}$ form factor $f_{+}$ significantly
contributes to reducing theoretical uncertainty on $|V_{us}|$ as well as the ratio of the decay constants $f_K$ and $f_{\pi}$.
The resulting theoretical uncertainty is now comparable with the level of precision of current experiments~\cite{Sachrajda:2011tg}.}.  A stringent test of CKM unitarity through 
the first row relation $|V_{ud}|^2+|V_{us}|^2+|V_{ub}|^2=1$ can be accomplished 
with the precision of $|V_{us}|$~\cite{BM}. 
A theoretical estimation of the vector coupling $f_1(0)$ is required to extract $V_{us}$ 
from the experimental rate of hyperon beta decay~\cite{{Cabibbo:2003cu},{Mateu:2005wi}}.

The matrix element for hyperon beta decays, $B_1\rightarrow B_2l\bar{\nu}$,
is composed of the vector and axial-vector transitions, $\langle B_2(p') | V_{\alpha}(x)+A_{\alpha}(x) | B_1(p) \rangle$,
which are described by six form factors: the vector ($f_1$), weak magnetism $(f_2)$, and induced scalar $(f_3)$ form factors
for the vector current, and the axial-vector $(g_1)$, weak electricity $(g_2)$, and induced pseudo-scalar $(g_3)$ form factors
for the axial current~\cite{Cabibbo:2003cu}. The experimental decay rate of the hyperon beta decay, $B_1\rightarrow B_2$, is given by
%
%
%
\bea
\Gamma&=&\frac{G_F^2}{60\pi^3}(M_{B_1}-M_{B_2})^5(1-3\delta)
|V_{us}|^2|f^{\Btob}_1(0)|^2 \cr
&&\times (1+\Delta_{\rm RC}) \left[1+3
\left|\frac{g^{\Btob}_1(0)}{f^{\Btob}_1(0)}\right|^2
+ \cdot\cdot\cdot
\right],
\label{Eq:DecayRate}
\eea
%
where $G_F$ is the Fermi constant measured from	the muon life time, 
which already includes some electroweak radiative corrections~\cite{Cabibbo:2003cu}.
The remaining radiative corrections to the decay rate are approximately represented by $\Delta_{\rm RC}$~\cite{Garcia:1985xz}.
Here, $M_{B_1}$ ($M_{B_2}$) denotes the rest mass of the initial (final) octet baryon state. 
The ellipsis can be expressed in terms of a power series in the small parameter 
$\delta=(M_{B_1}-M_{B_2})/(M_{B_1}+M_{B_2})$, which is regarded as a size of flavor SU(3) breaking~\cite{Gaillard:1984ny}.
The first linear term in $\delta$, which should be given by 
$-4\delta [g_2(0)g_1(0)/f_1(0)^2]_{\Btob}$~\footnote{
Conventionally, $(M_{B_1}-M_{B_2})/M_{B_1}$ is adopted in Eq.~(\ref{Eq:DecayRate}) to 
be the small parameter $\delta$~\cite{{Gaillard:1984ny},{Cabibbo:2003cu}}.
However, our definition of the SU(3) breaking parameter, $\delta=(M_{B_1}-M_{B_2})/(M_{B_1}+M_{B_2})$, is theoretically preferable 
for considering the time-reversal symmetry on the matrix elements of hyperon beta decays in lattice 
QCD calculations~\cite{{Guadagnoli:2006gj},{Sasaki:2008ha}}. Accordingly, a factor of $(M_{B_1}+M_{B_2})/M_{B_1}$ is different 
in definitions of $g_2$, $g_3$, $f_2$ and $f_3$ form factors in comparison to those adopted in experiments.
},
is safely ignored as small as ${\cal O}(\delta^2)$ since the nonzero value of 
the second-class form factor $g_2$~\cite{Weinberg:1958ut} should be induced 
at first order of the $\delta$ expansion~\cite{Gaillard:1984ny}. 
The absolute value of $g_1(0)/f_1(0)$ can be determined
by measured asymmetries such as electron-neutrino 
correlation~\cite{{Cabibbo:2003cu},{Gaillard:1984ny}}.
A theoretical attempt to evaluate SU(3)-breaking corrections on the vector coupling $f_1(0)$, whose value
is given by SU(3) Clebsch-Gordan coefficients in the exact SU(3) limit, is primarily required for the precise 
determination of $|V_{us}|$.

The value of $f_1(0)$ should be equal to the SU(3) Clebsch-Gordan coefficients up to the second 
order in SU(3) breaking, thanks to the Ademollo-Gatto theorem (AGT)~\cite{Ademollo:1964sr}. 
As the mass splittings among octet baryons are typically of the order of 10-15\%, an expected size of the 
second-order corrections is a few percent level. However, either the size or the sign of their corrections is 
somewhat controversial among various theoretical studies at present as summarized in Table~\ref{Tab:Th_estimate_f_1}. 
{\it A model independent evaluation} of SU(3)-breaking corrections is highly desired. 
Although recent quenched lattice studies suggest that the second-order correction on $f_1(0)$ is likely 
negative~\cite{{Guadagnoli:2006gj},{Sasaki:2008ha}}, 
we need further confirmation from (2+1)-flavor dynamical lattice QCD near the physical point.

Our paper is organized as follows. In Sec. II, we first summarize the numerical
lattice QCD ensembles used for this work and then give the details of
our Monte Carlo simulations. The numerical results are presented in Sec. III.
We begin with our determination of the scalar form factor $f_S(q^2)$, which will be defined in the later session,
at finite momentum transfer. We discuss in detail the interpolation of the form factor to zero momentum transfer and
also the chiral extrapolation of the hyperon vector coupling $f_1(0)$. Finally, in Sec. VI, 
we summarize our results and conclusions.

\begin{table*}[tbp]
\begin{center}
\caption{Theoretical uncertainties of 
$\tilde{f}_1(0)=f_1(0)/f_1^{\rm SU(3)}(0)$ for various hyperon beta-decays.
HBChPT and EOMS-CBChPT stand for heavy baryon chiral perturbation theory and
covariant baryon chiral perturbation theory with
the extended on-mass-shell (EOMS) renormalization scheme.
\label{Tab:Th_estimate_f_1}
}
\begin{ruledtabular}
\begin{tabular}{|  l  |  l l l l     |}
\hline
Type of result (reference)
&$\Lambda\rightarrow p$ & $\Sigma^- \rightarrow n$ & $\Xi^- \rightarrow \Lambda$ & $\Xi^0
\rightarrow\Sigma^+$ \\
\hline
Bag model~\cite{Donoghue:1981uk}  & 0.97 & 0.97 & 0.97 & 0.97 \\
Quark model~ \cite{Donoghue:1986th} & 0.987 & 0.987 & 0.987 & 0.987 \\
Quark model~\cite{Schlumpf:1994fb} &0.976&0.975&0.976&0.976\\
$1/N_c$ expansion~\cite{Flores-Mendieta:1998ii}
  &1.02(2)&1.04(2)&1.10(4)&1.12(5)\\
Full ${\cal O}(p^4)$ HBChPT~\cite{Villadoro:2006nj} &1.027&1.041&1.043&1.009\\
Full ${\cal O}(p^4)$ + partial ${\cal O}(p^5)$ HBChPT~\cite{Lacour:2007wm} 
& 1.066(32) & 1.064(6) & 1.053(22) & 1.044(26) \\
Full ${\cal O}(p^4)$ EOMS-CBChPT~\cite{Geng:2009ik} &0.943(21) &1.028(02)& 0.989(17) & 0.944(16)\\
Full ${\cal O}(p^4)$ EOMS-CBChPT + Decuplet~\cite{Geng:2009ik} 
&1.001(13) &1.087(42)& 1.040(28) & 1.017(22)\\
\hline
Quenched lattice QCD~\cite{{Guadagnoli:2006gj},{Sasaki:2008ha}} & N/A & 0.988(29) & N/A & 0.987(19)\cr
\hline
\end{tabular}
\end{ruledtabular}
\end{center}
\end{table*}

\section{Simulation details}

In this paper, we will present the first result 
for the hyperon vector form factor for 
$\Xi^0\rightarrow \Sigma^+ l\bar{\nu}$ and $\Sigma^-\rightarrow n l\bar{\nu}$ semileptonic decays 
from simulations with 2+1 flavors of domain wall fermions (DWFs).
We use the RBC and UKQCD collaboration ensembles, 
which are generated on a $24^3\times 64$ lattice with two light degenerate quarks and a single flavor heavier quark and
the Iwasaki gauge action at $\beta=2.13$~\cite{Allton:2008pn}. The dynamical light and strange quarks are described
by DWF actions with fifth dimensional extent $L_s=16$
and the domain-wall height of $M_5=1.8$, which give
a residual mass of $am_{\rm res}\approx 0.003$. 
Each ensemble of configurations uses the same 
dynamical strange quark mass, $am_{s}=0.04$, which is close to 
its physical value~\cite{Allton:2008pn}. We have already published our findings 
in nucleon structure from the same ensembles 
in three publications, Refs.~\cite{{Yamazaki:2008py},{Yamazaki:2009zq},{Aoki:2010xg}}.

The inverse of lattice spacing is $a^{-1}=1.73(3)$ [$a$=0.114(2) fm], 
which is determined from the $\Omega^{-}$ baryon mass~\cite{Allton:2008pn}. 
Accordingly, the physical spatial extent is approximately 2.7 fm, where
the nucleon vector form factor at low $q^2$ doesn't suffer much 
from the finite size effect though such effect may influence
other nucleon form factors~\cite{{Yamazaki:2008py},{Yamazaki:2009zq}}.
We choose three values for the light quark masses, 
$am_{ud}=0.005$, 0.01, and 0.02, which correspond to about
330 MeV, 420 MeV and 560 MeV pion masses~\footnote{
Preliminary results obtained at $am_{ud}=0.005$ were first reported in Ref.~\cite{Sasaki:2011hu}.}. 
We use 4780, 2350, and 1580 trajectories separated by 20 trajectories for $am_{ud}=0.005$,  0.01, 
and 0.02~\cite{Allton:2008pn}. The total number of configurations is 240 for $am_{ud}=0.005$, 120 for $am_{ud}=0.01$,
and 80 for $am_{ud}=0.02$ as summarized in Table~\ref{Tab:Summary_para}.

We make four (two) measurements on each configuration using a single source location, 
which is located at $(x, y, z, t)= (6n, 6n, 6n, 16n)$
with $n=0, 1, 2, 3$ ($n=0, 2$) for $am_{ud}=0.02$ ($am_{ud}=0.01$ and $0.005$), and then they are averaged on each configuration in order to
reduce possible autocorrelations among measurements. The statistical errors are
estimated by the jackknife method on such blocked measurements.
The quark propagators are calculated by gauge-invariant Gaussian smearing at the source with
smearing parameters $(N,\omega)=(100, 7)$. 
Details of our calculation of the quark propagators are described in Ref.~\cite{Yamazaki:2009zq}.

\begin{table*}[ht]
\caption{
$N_{\rm conf}$, $N_{\rm sep}$, and $N_{\rm meas}$ denote the number
fo gauge configurations, trajectory separation between each measured
configuration, and the number of measurements on each configuration, respectively.
The table contains the pion, kaon, nucleon, $\Sigma$-baryon, and $\Xi$-baryon mass for each ensemble.
}\label{Tab:Summary_para}
\begin{ruledtabular}
\begin{tabular}{cccclllll}
\hline
$m_{ud}$ & $N_{\rm conf}$ & $N_{\rm sep}$ & $N_{\rm meas}$ & $M_{\pi}$ [GeV]
& $M_{K}$ [GeV] & $M_{N}$ [GeV] & $M_{\Sigma}$ [GeV]  & $M_{\Xi}$ [GeV]  \cr
\hline
0.005 & 240 & 20 & 2 & 0.3297(7) & 0.5759(8)  & 1.140(12) & 1.330(9) & 1.431(6)\cr
0.01 &120 & 20 & 2 & 0.4200(12) & 0.6064(11)  &  1.237(13) & 1.386(12) & 1.465(8)\cr
0.02 &80 & 20 & 4 & 0.5580(11) & 0.6651(11) & 1.412(10) & 1.501(9) & 1.544(8)\cr
\hline 
\hline
\end{tabular}
\end{ruledtabular}
\end{table*}

\section{Numerical Results}

\subsection{Scalar form factor $f_S(q^2)$ at $q^2=q_{\rm max}^2$}
We focus on vector couplings $f_1(0)$ for two different hyperon beta-decays, 
$\Xi^0\rightarrow \Sigma^+ l\bar{\nu}$ and $\Sigma^-\rightarrow n l\bar{\nu}$. These decays are 
simply denoted by $\XtoS$ and $\StoN$ hereafter. We recall that 
$f^{\XtoS}_1(0)=+1$ and $f^{\StoN}_1(0)=-1$ in the exact SU(3) 
limit. For convenience in numerical calculations, instead of the vector form factor $f_1(q^2)$, 
we consider the so-called scalar form factor
\be
f^{\Btob}_S(q^2)=f^{\Btob}_1(q^2)+\frac{q^2}{M_{B_1}^2-M_{B_2}^2}f^{\Btob}_3(q^2),
\ee
where $f_3$ represents the second-class form factor, which is identically zero in the exact SU(3) limit~\cite{Weinberg:1958ut}. 
The renormalized value of $f_S(q^2)$ at $q_{\rm max}^2=-(M_{B_1}-M_{B_2})^2 <0$~\footnote{We note that
$q^2$ quoted here is defined in the Euclidean metric convention. See details of our convention found in Ref.~\cite{Sasaki:2008ha}.} can be precisely 
evaluated by the double ratio method proposed in Ref.~\cite{Guadagnoli:2006gj},
where all relevant three-point functions are determined at zero three-momentum transfer $|{\bf q}|=0$.
For the three-point functions, we use the sequential source method. 
We use the source-sink separation of 12 lattice units following previous works related to nucleon structure
\cite{{Yamazaki:2008py},{Yamazaki:2009zq},{Aoki:2010xg}}.
Details of the construction of the three-point functions from the sequential quark propagator 
are described in Ref.~\cite{Sasaki:2003jh}.

Here we note that the absolute value of 
the renormalized $f_S(q_{\rm max}^2)$ is exactly unity in the flavor SU(3) symmetric limit, where $f_S(q^2_{\rm max})$ becomes $f_1(0)$, for the hyperon decays considered here. Thus, the
deviation from unity in $|f_S(q^2_{\rm max})|$ is attributed to three types of the SU(3) breaking effect:
(1) the recoil correction ($q^2_{\rm max}\neq 0$) stemming from the mass difference of $B_1$ and $B_2$ states,
(2) the presence of the second-class form factor $f_3(q^2)$, and 
(3) the deviation from unity in the renormalized $f_1(0)$. Taking the limit of zero four-momentum transfer
of $f_S(q^2)$ can separate the third effect from the others, since the scalar form factor at $q^2=0$, $f_S(0)$,
is identical to $f_1(0)$. Indeed, our main target is to measure the third one.

\begin{figure}
\begin{center}
\includegraphics[width=1\columnwidth,clip]{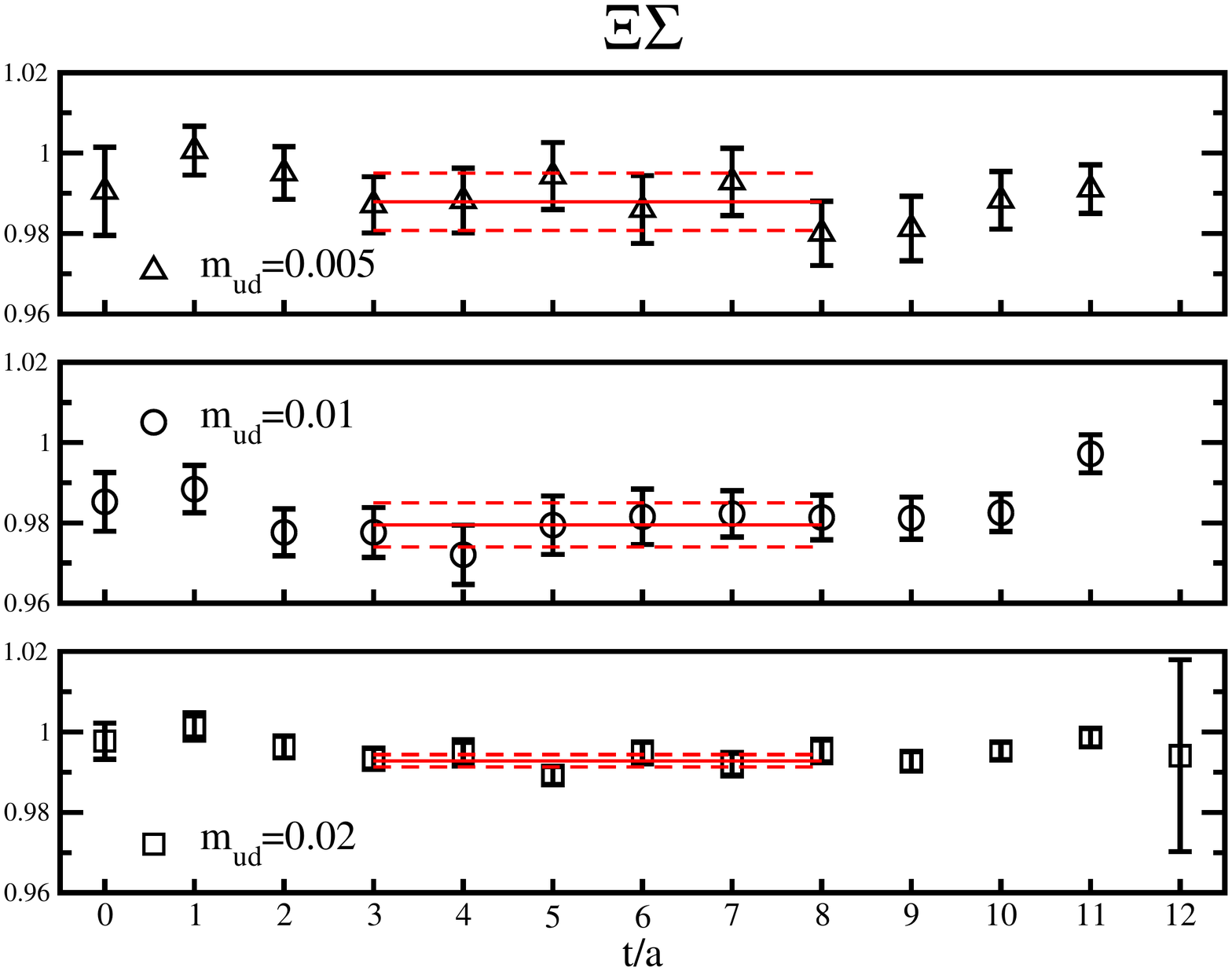}
\includegraphics[width=1\columnwidth,clip]{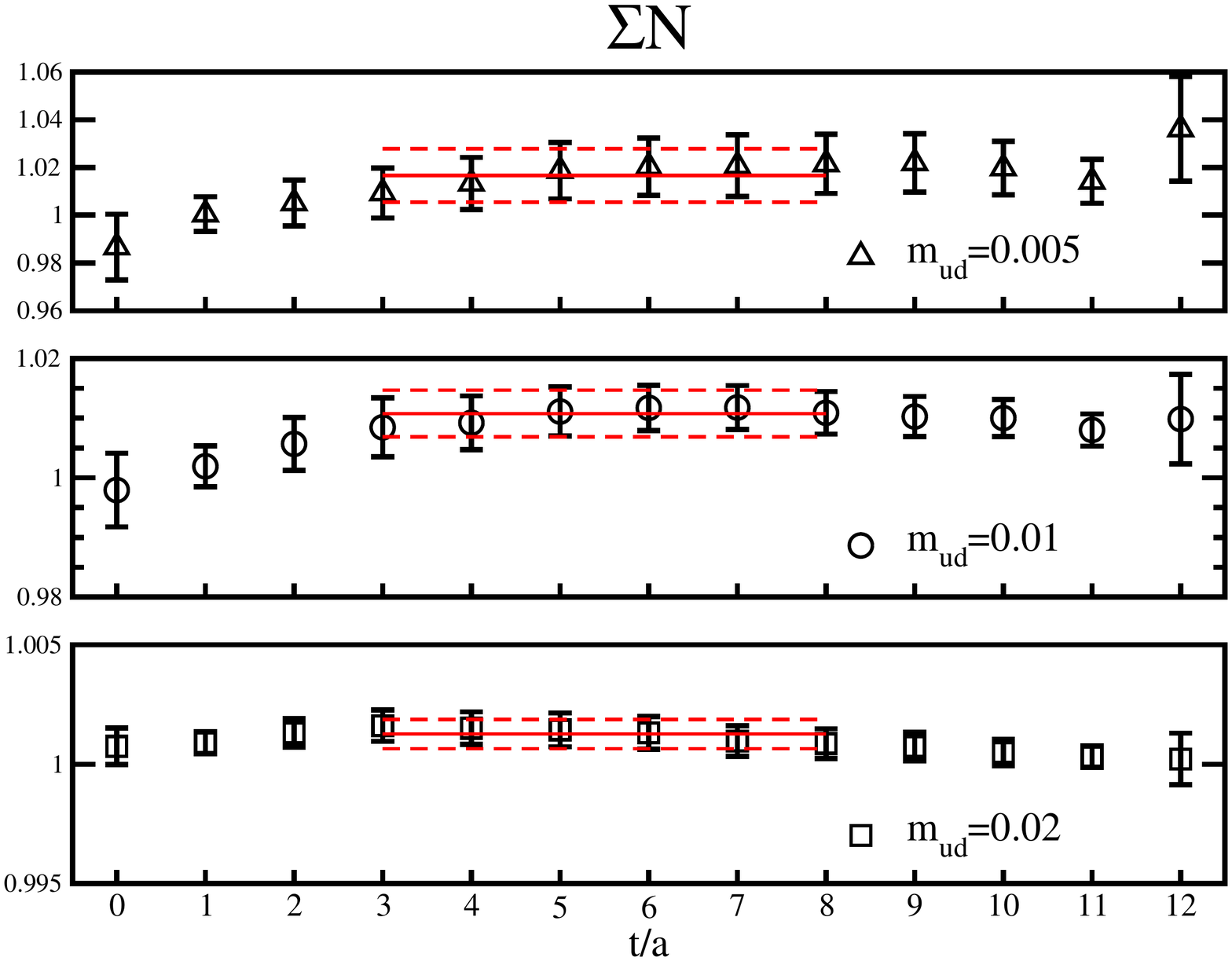}
\end{center}
\caption{The absolute value of $f_{S}^{\rm ren}(q_{\rm max}^2)$ as a function of the current insertion time slice.
The upper (lower) panel is for $\XtoS$ ($\StoN$) decay.
In each panel, results for $am_{ud}=0.005$, 0.01, and 0.02 are plotted from top to bottom.
The	lines	represent	the average value (solid lines) and their 1 standard deviations (dashed lines)
over range of $3\le t/a\le 8$.
}
\label{Fig:F0max}
\end{figure}

In Fig.~\ref{Fig:F0max}, we plot the absolute value of the renormalized $f_{S}(q_{\rm max}^2)$ as a function of the
current insertion time slice. Good plateaus are observed in the middle region
between the source and sink points. The lines represent the average value (solid lines)
and their 1 standard deviations (dashed lines) over range of $3\le t/a\le 8$. 
The obtained values of $|f_S(q_{\rm max}^2)|$, which are naturally renormalized in the double ratio method, as well as $q_{\rm max}^2$ values
are summarized in Table~\ref{Tab:atqmax}.

\subsection{Interpolation to zero four-momentum squared}
The scalar form factor $f_S(q^2)$ at $q^2>0$ is also calculable 
with nonzero three-momentum transfer ($|{\bf q}|\neq 0$).
To avoid unnecessary repetition, we simply give a reference~\cite{Sasaki:2008ha},
where all the technical details are available.

We use the four lowest nonzero momenta: ${\bf q}=2\pi/L\times(1,0,0)$,
$(1,1,0)$, $(1,1,1)$, and $(2,0,0)$, corresponding to a $q^2$ range
from about 0.2 to 0.8 GeV$^{2}$.
We then can make the $q^2$ interpolation of $f_S(q^2)$ to $q^2=0$
by the values of $f_S(q^2)$ at $q^2>0$
together with the precisely measured value of $f_S(q^2)$ at $q^2=q^2_{\rm max}<0$
from the double ratio. 

\begin{figure*}
\begin{center}
\includegraphics[width=0.65\columnwidth,clip]{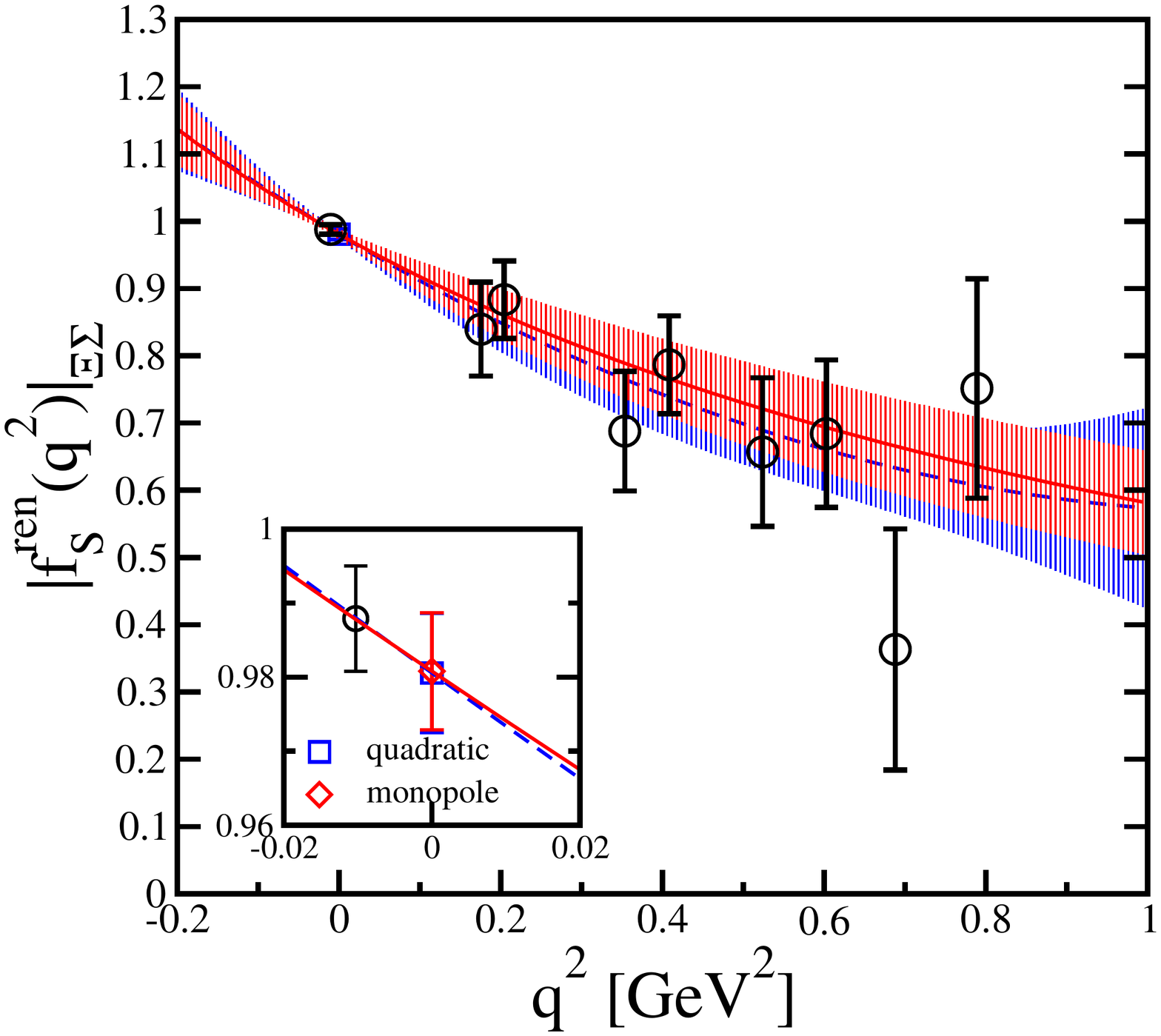}
\includegraphics[width=0.65\columnwidth,clip]{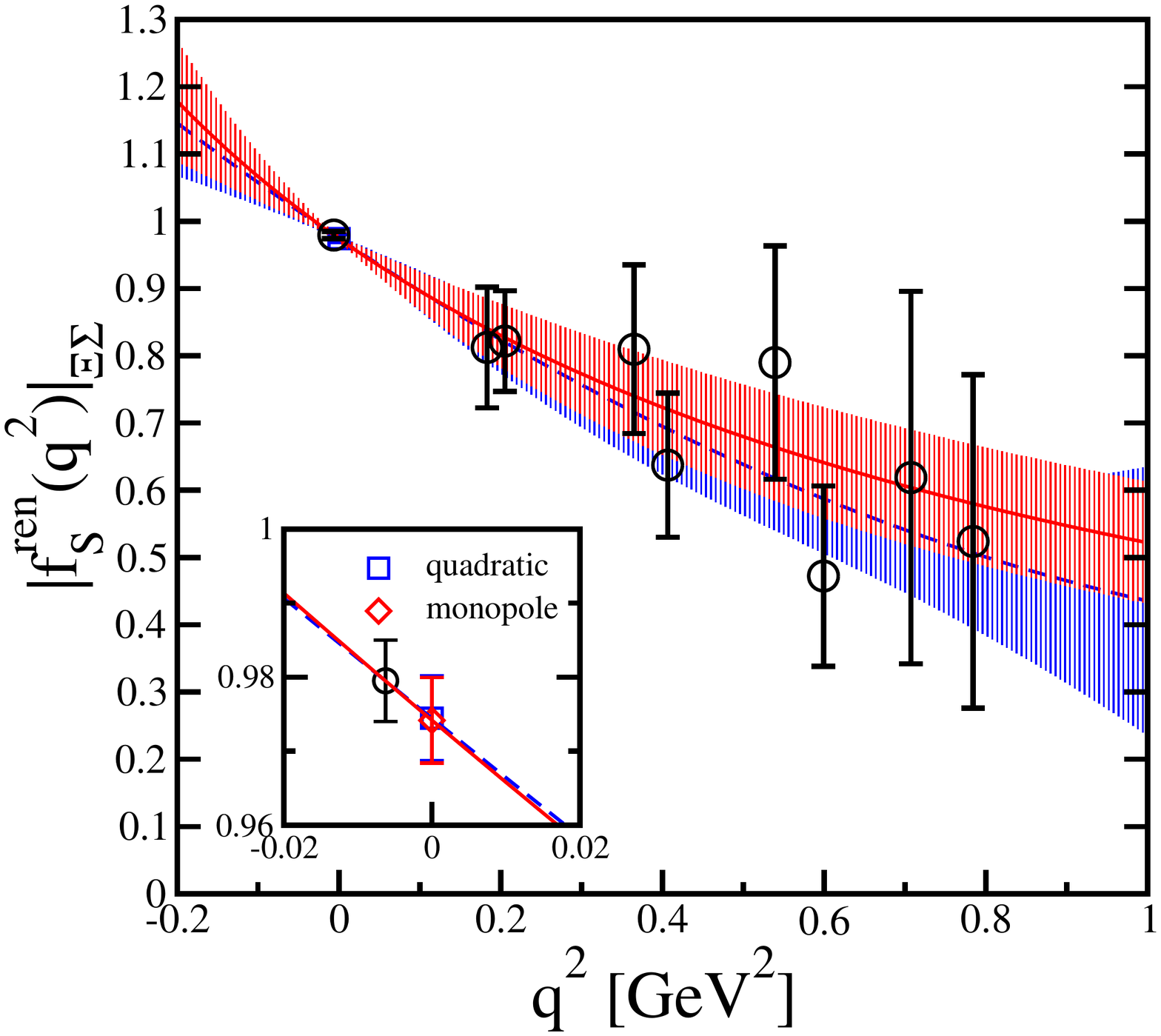}
\includegraphics[width=0.65\columnwidth,clip]{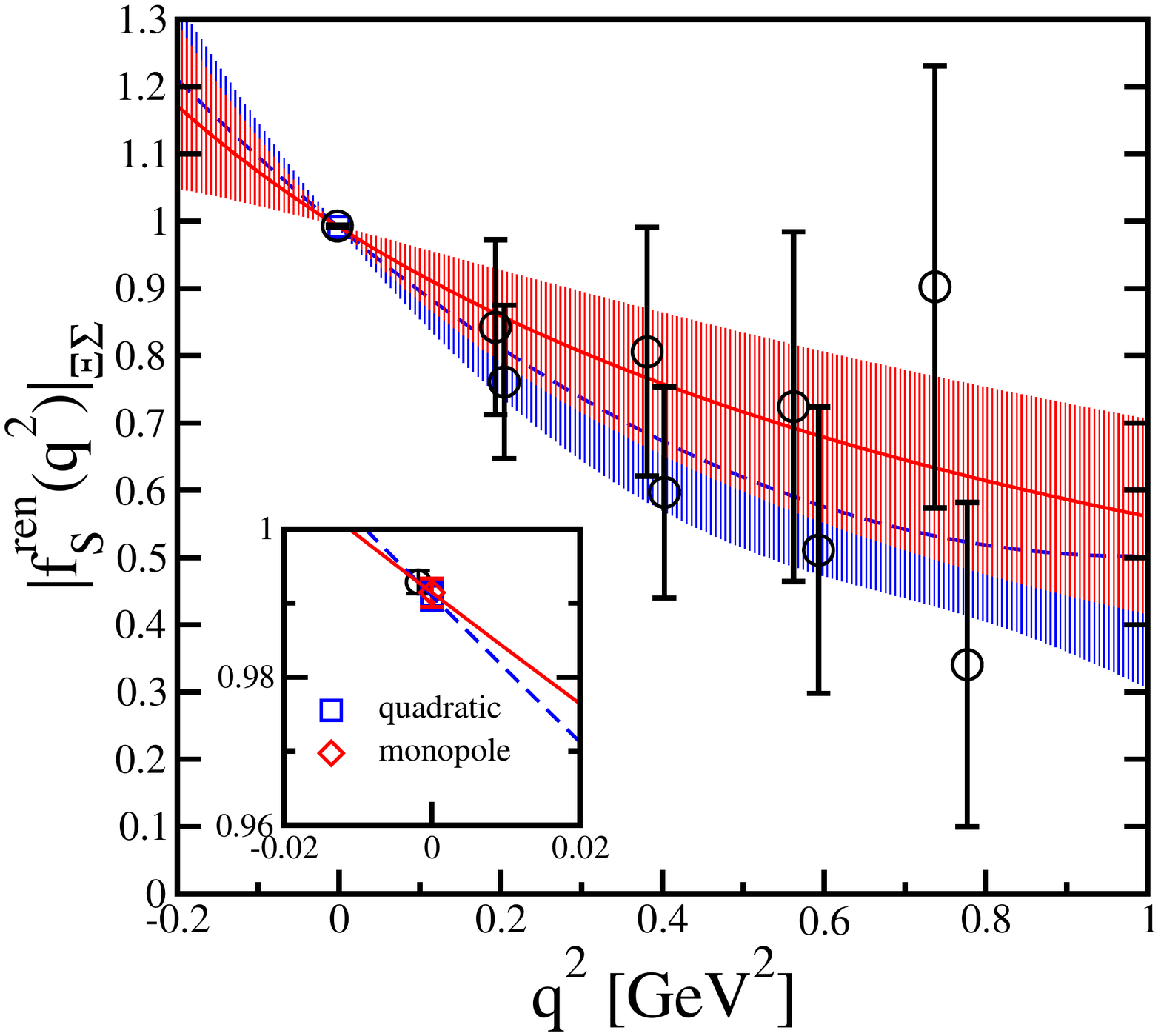}
\includegraphics[width=0.65\columnwidth,clip]{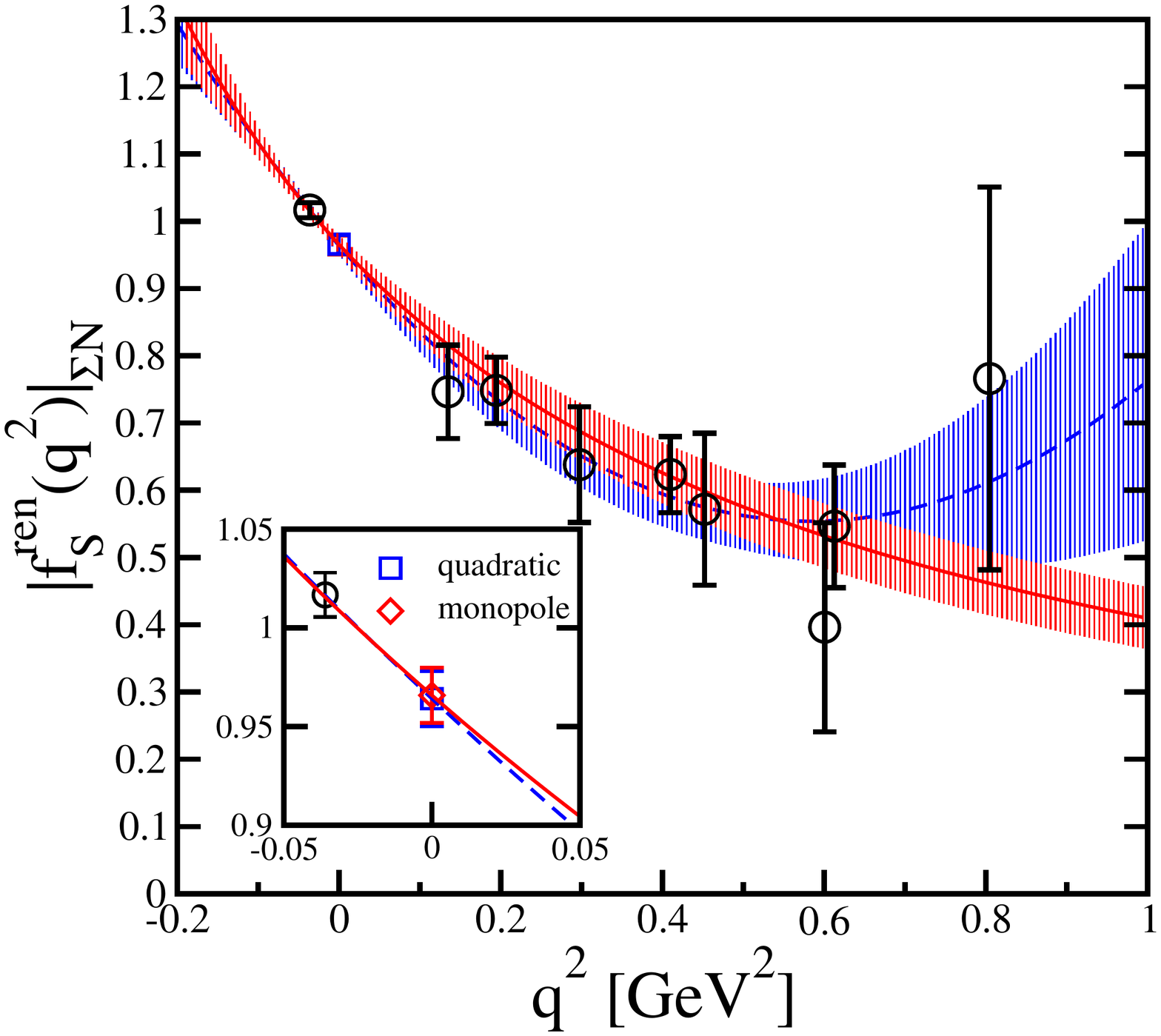}
\includegraphics[width=0.65\columnwidth,clip]{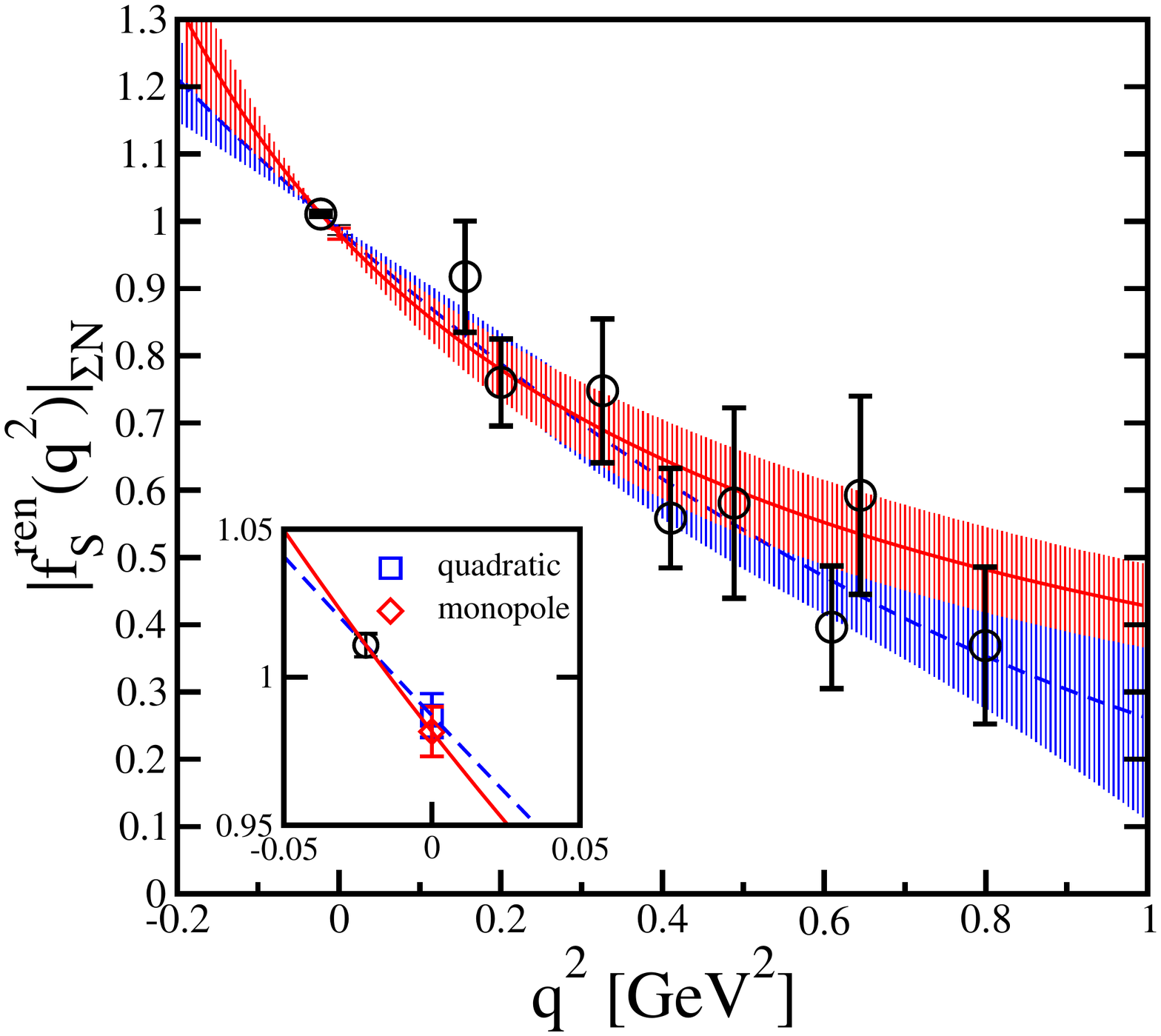}
\includegraphics[width=0.65\columnwidth,clip]{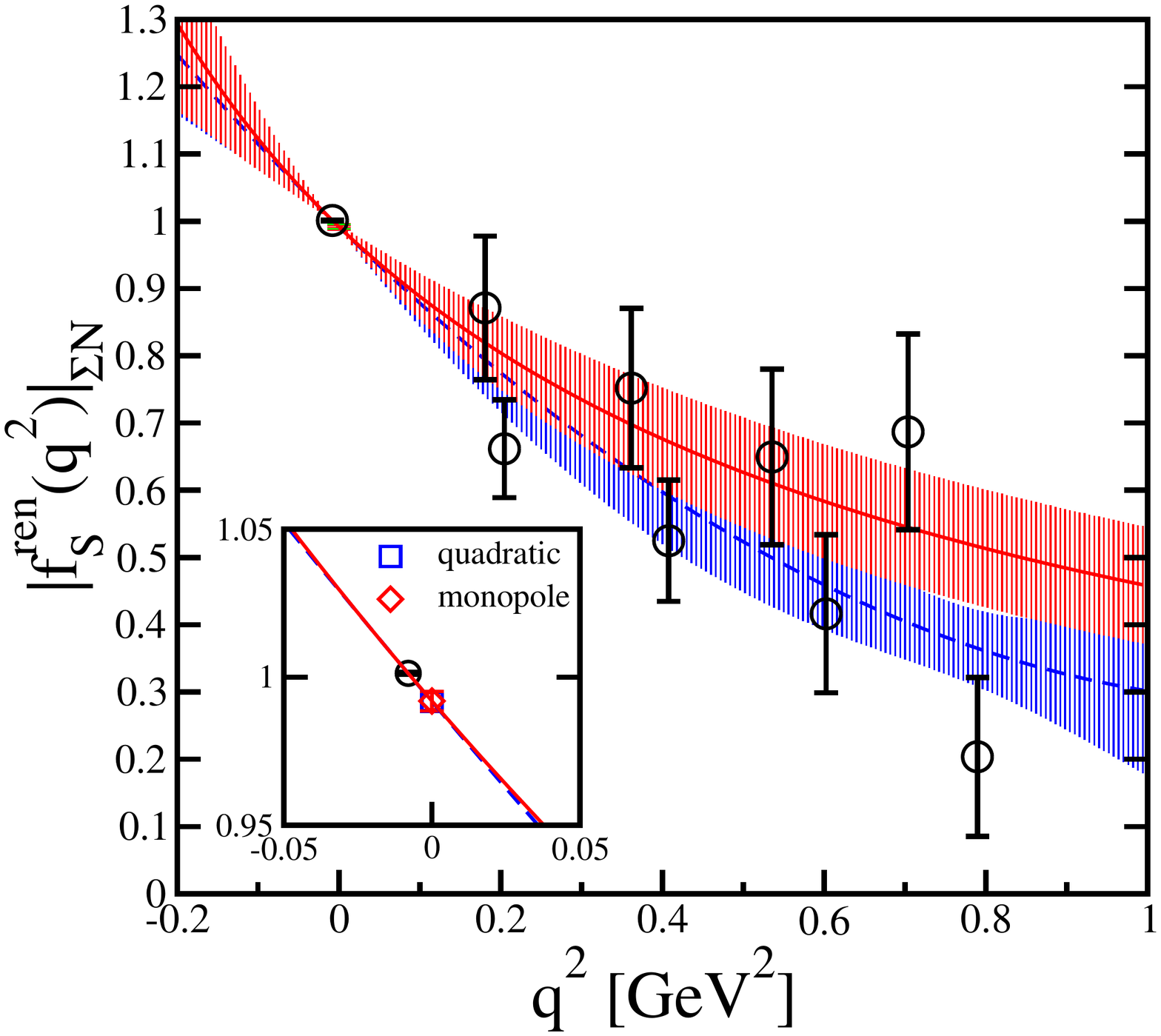}
\end{center}
\caption{Interpolation of $f_S(q^2)$ to $q^2=0$.
The upper (lower) panels are for $\XtoS$ ($\StoN$) decay at $am_{ud}=0.005$ (left), 0.01 (middle), and 0.02 (right). 
Open circles are $|f_S(q^2)|$ at the simulated $q^2$. The solid (dashed) curve is the fitting
result by using monopole (quadratic) interpolation form, while the open diamond (square) represents
the interpolated value to $q^2=0$.
}
\label{Fig:F0_Extra_q2}
\end{figure*}

In Fig.~\ref{Fig:F0_Extra_q2}, we plot the absolute value of 
the renormalized $f_S(q^2)$ as a function of $q^2$ for $\XtoS$ (upper panels) 
and $\StoN$ (lower panels) at $am_{ud}=0.005$ (left), 0.01 (middle) and 0.02 (right).
In this work, we also calculate the time-reversal process $\btoB$ as well as $\Btob$, to get more
data points in the $q^2>0$ region.
Open circles are $|f_S(q^2)|$ at the simulated $q^2$. The solid (dashed)
curve is the fitting result with the seven lowest-$q^2$ data points by using the monopole (quadratic) interpolation
form~\cite{Sasaki:2008ha}, while the open diamond (square) represents the interpolated value
to $q^2=0$. 

As shown in Fig.~\ref{Fig:F0_Extra_q2}, two determinations to
evaluate $f_S(0)=f_1(0)$ from measured points are indeed consistent
with each other. Thus, this observation indicates that the choice of the interpolation form does not affect the interpolated
value $f_1(0)$ significantly. We simply prefer to use the values obtained from the monopole fit
in the following discussion.

\subsection{Chiral extrapolation of $f_1(0)$}
\begin{figure*}
\begin{center}
\includegraphics[width=0.9\columnwidth,clip]{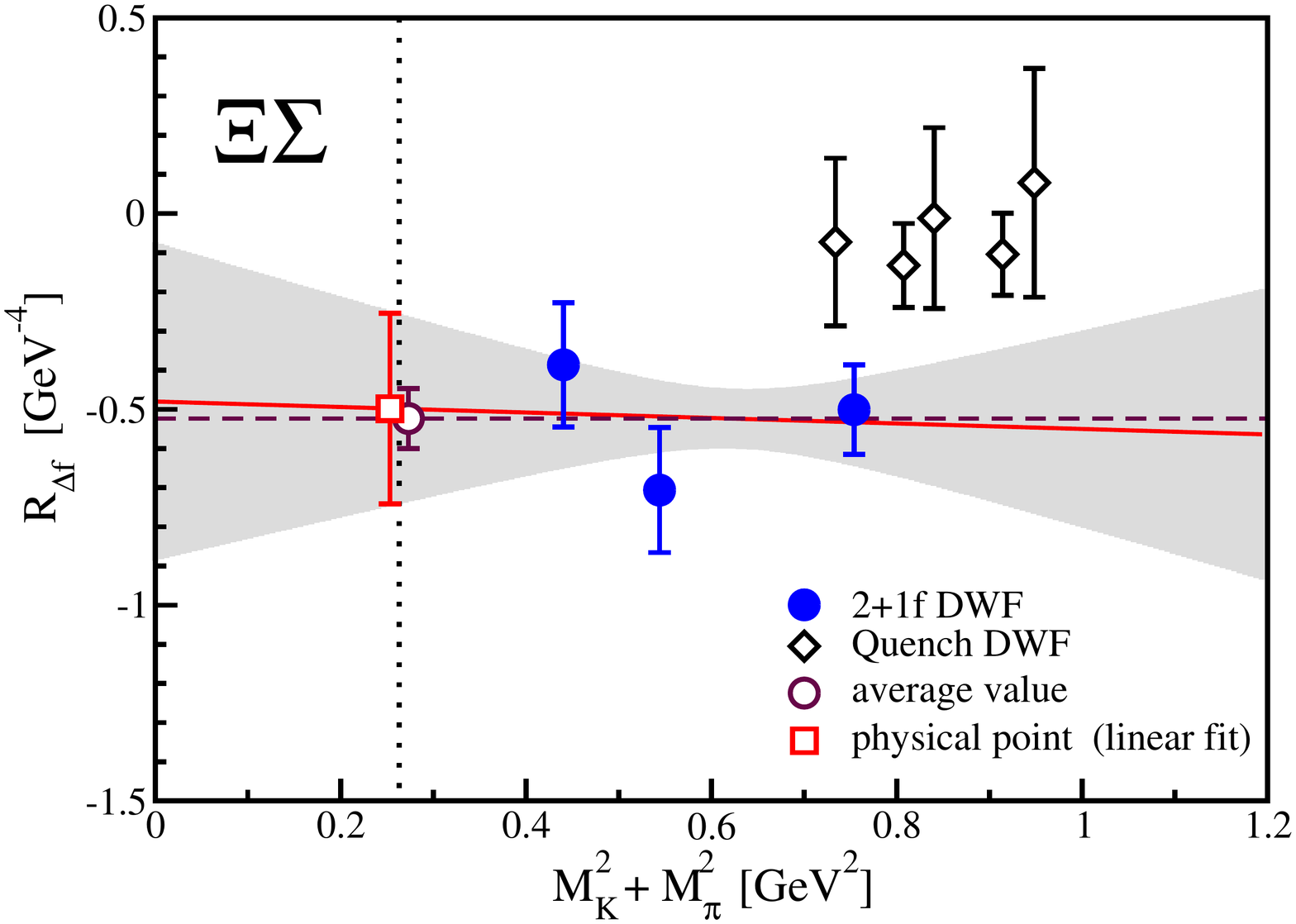}
\includegraphics[width=0.9\columnwidth,clip]{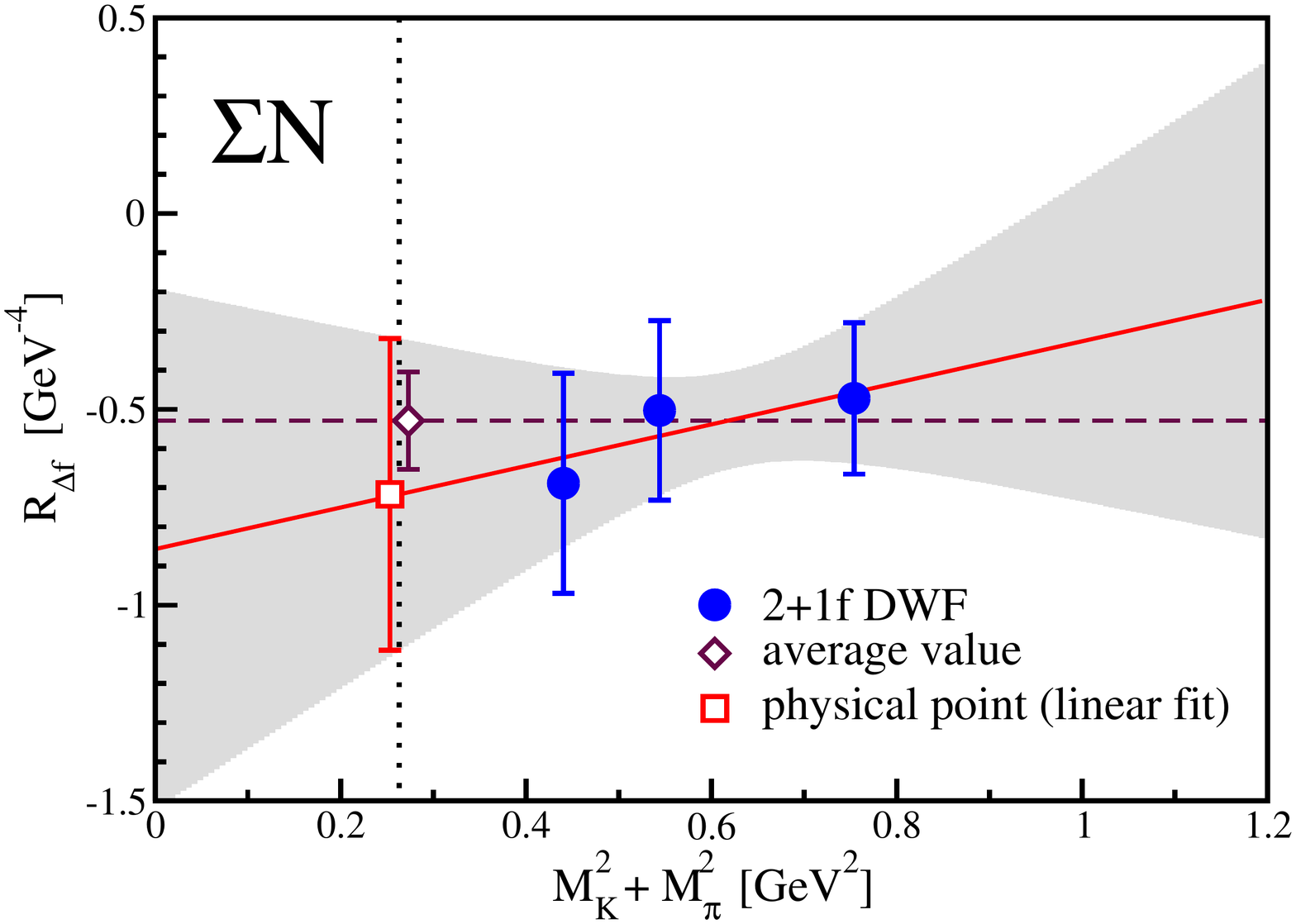}
\end{center}
\caption{Chiral extrapolation of $R_{\Delta f}$ for $\XtoS$ (left) and $\StoN$ (right).
In each panel, solid and dashed lines represent the fitting results obtained by a linear and
a constant fit, respectively. The dotted vertical lines mark the position of the physical point.
The open square (circle) symbols, which have been shifted slightly to the left (right), represent
the extrapolated values at the physical point.
In the right panel, quenched DWF results~\cite{Sasaki:2008ha} are also included 
as open diamonds for a comparison.}
\label{Fig:DRF_Extra_Phys}
\end{figure*}

In order to estimate $f_1(0)$ at the physical point, we perform the chiral extrapolation of $f_1(0)$. 
The ratio of $f_{1}(0)/f_1^{\rm SU(3)}$ can be parametrized as 
$\tilde{f}_1(0)=f_{1}(0)/f_1^{\rm SU(3)}(0)=1+\Delta f$, where $\Delta f$ represents all SU(3)-breaking 
corrections on $f_1(0)$.
We then introduce the following ratio~\cite{{Guadagnoli:2006gj},{Sasaki:2008ha}}:
\be
R_{\Delta f}(M_K, M_\pi)=\frac{\Delta f}{(M_K^2 - M_{\pi}^2)^2},
\label{Eq:AGTform}
\ee
where the leading symmetry-breaking correction, which is predicted by the Ademollo-Gatto
theorem, is explicitly factorized out. The remaining dependence related to either the higher order corrections of the SU(3) breaking 
or simulated pion and kaon masses is hardly observed within the statistical errors as 
shown in Fig.~\ref{Fig:DRF_Extra_Phys}.

Indeed, if we simply adopt a linear fit form on $\Delta R$ as 
a function of $M_{K}^2+M_{\pi}^2$ to extrapolate the value at the physical point:
\be
R_{\Delta f}(M_K, M_\pi)=R_0 + R_1 \cdot (M_{K}^2+M_{\pi}^2),
\ee
the resulting coefficient $R_1$, which is approximately zero, ensures that the remaining dependence of either $M_K$ or $M_\pi$
is negligible at least within the current statistics. 
This observation suggests that although the simulated strange quark mass is slightly heavier than the physical mass,
the corresponding systematic error is likely to be small in the chiral extrapolation of $R_{\Delta f}$.

We may rather use fitting the data of $R_{\Delta f}$ to a constant to estimate the value at the physical point.
The two fits are mutually consistent, but the latter provides the smaller error
as shown in Fig.~\ref{Fig:DRF_Extra_Phys}. All fitted results are also tabulated in Table~\ref{Tab:R_Df}
as well as the values of $R_{\Delta f}$ given at all simulated quark masses. We thus quote the value of $R_{\Delta f}$ 
at the physical point:
\be
R_{\Delta f}(M_K^{\rm phys}, M_{\pi}^{\rm phys})=\left\{
\begin{array}{ll}
-0.524(77) & {\rm for}\;\XtoS\cr
-0.529(125) & {\rm for}\;\StoN,
\end{array}
\right.
\ee
in $({\rm GeV})^{-4}$, which is obtained from the latter fit, as our best estimate. 
We evaluate the SU(3)-breaking correction $\Delta f$ via Eq.~(\ref{Eq:AGTform}) 
together with the physical kaon and pion masses and then get
\be
\tilde{f}^{\rm AGT}_{1}(0)=1+\Delta f = \left\{
\begin{array}{ll}
0.9737(39) & {\rm for}\;\XtoS\cr
0.9734(63) & {\rm for}\;\StoN,
\end{array}
\right.
\ee
which we call the AGT fit result hereafter. Alternatively, we may perform a global fit of
the data on $\tilde{f}_{1}(0)$ as multiple functions of $M_{K}^2-M_{\pi}^2$ and 
$M_{K}^2+M_{\pi}^2 $
\be
\tilde{f}_{1}(0)=C_0 + \left( C_1 +  C_2 \cdot (M_{K}^2+M_{\pi}^2 )\right)\cdot (M_{K}^2-M_{\pi}^2)^2 ,
\label{Eq:GlobalFit}
\ee
whose form is motivated by the AGT fit. Our simulations are performed
with a strange quark mass slightly heavier than the physical mass.
To take into account this slight deviation in this global analysis of the chiral extrapolation, we simply
evaluate a correction using the Gell-Mann-Oakes-Renner relation for the pion and kaon masses,
which corresponds to the quark mass dependence of pseudo-scalar meson masses at the
leading order of ChPT. This correction could be accurate in as much as the ratio of 
$R_{\Delta f}$ have shown neither any higher-order corrections of SU(3) breaking nor
the remaining $M_K$ and $M_\pi$ dependences. 

\begin{figure*}
\begin{center}
\includegraphics[width=0.9\columnwidth,clip]{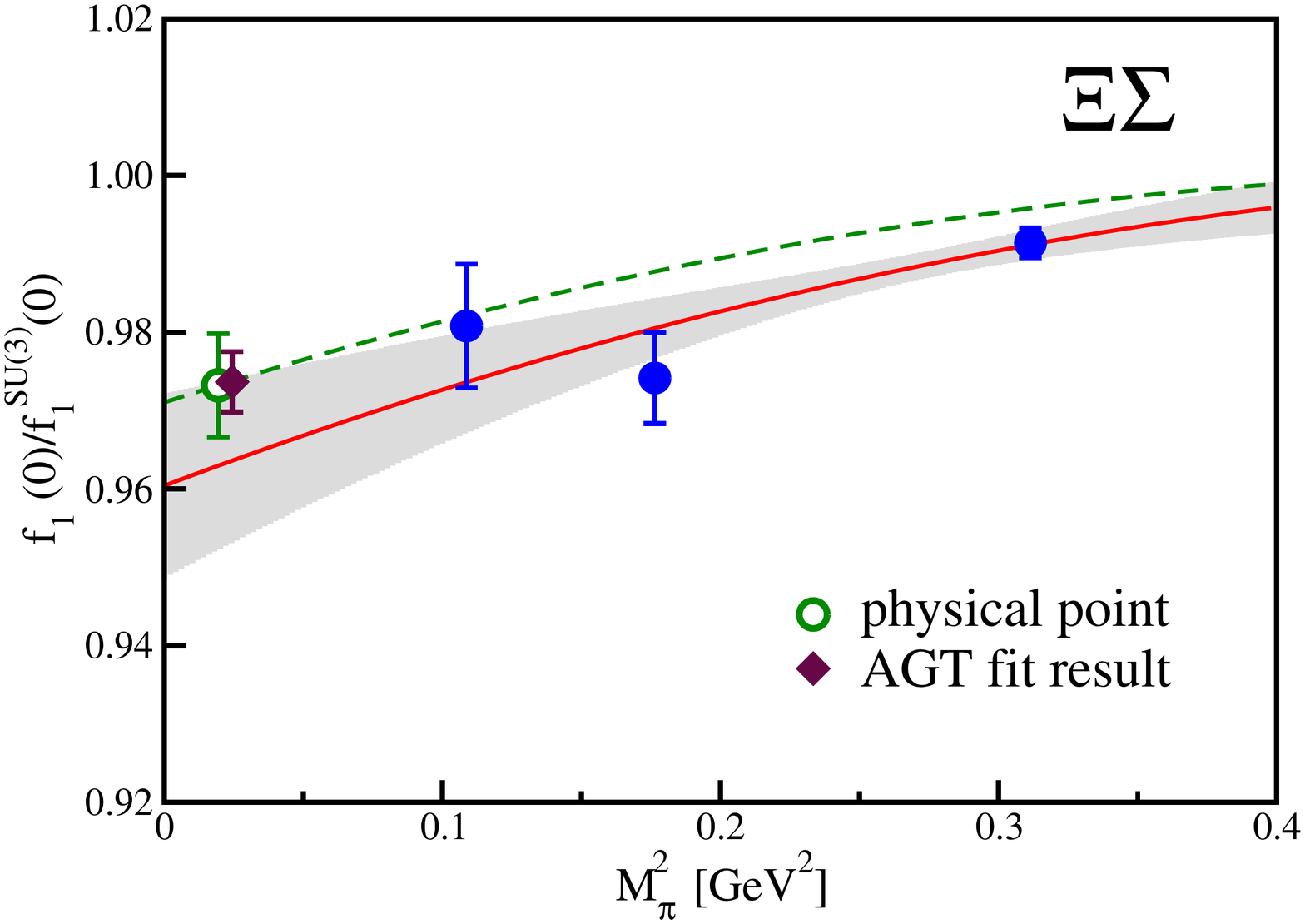}
\includegraphics[width=0.9\columnwidth,clip]{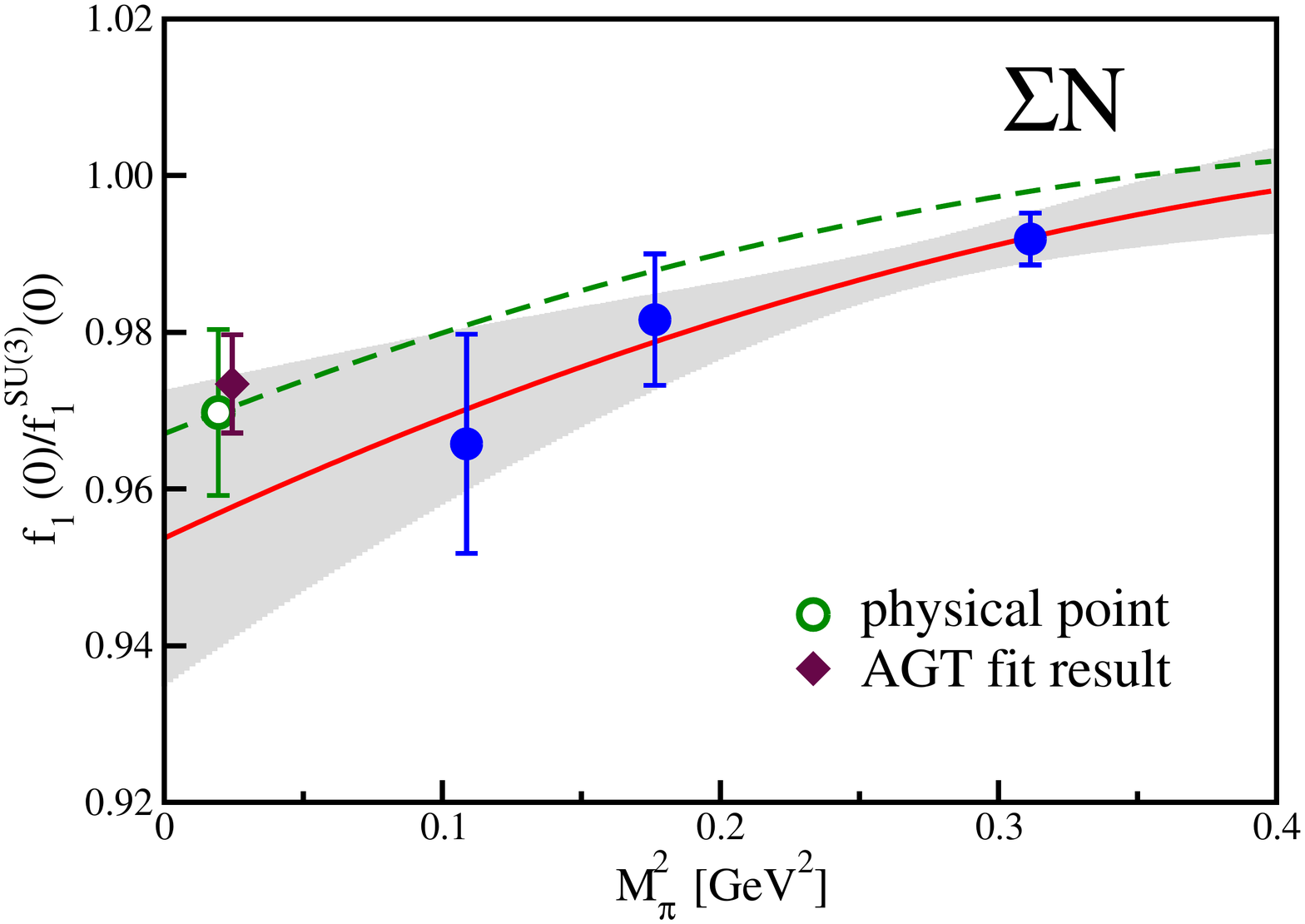}
\end{center}
\caption{Chiral extrapolation of $\tilde{f}_1(0)$ for $\XtoS$ (left) and $\StoN$ (right).
In each panel, the filled circles denote our results obtained with the three ensembles, while 
open circles are extrapolated results at the physical point using a global fitting procedure described in the text.
Fitting curves indicated by dashed and solid curves represent the fitting results for the physical strange mass
and the simulated one ($am_s=0.04$), respectively.
The AGT fit results are also included as filled diamond symbols, which have been moved slightly 
to the right, for comparison.
}
\label{Fig:F1_Extra_Phys}
\end{figure*}

In Fig.~\ref{Fig:F1_Extra_Phys}, 
we present the results of $\tilde{f}_1(0)$ (filled circles) as a function of the pion mass squared
for $\XtoS$ (left panel) and $\StoN$ (right panel).
In each panel, fitting curves indicated by dashed and solid curves represent the fitting results 
with and without the correction for the strange quark mass, respectively. 
The extrapolated results of $\tilde{f}_1(0)$ at the physical point, which are denoted as open circles, 
agree very well with the AGT fit results indicated by filled diamond symbols.
Both results are tabulated in Table~\ref{Tab:FinalResults} together with the data calculated
at all simulated quark masses. The statistical errors from the AGT fit are rather smaller than
those of the global fits.

\begin{table}[ht]
\caption{
Results for $|f^{\rm ren}_S(q_{\rm max}^2)|$, where $q_{\rm max}^2=-(M_{B_1}-M_{B_2})^2$
with $(B_1,B_2)=(\Xi,\Sigma)$ and $(\Sigma, N)$.
}
\label{Tab:atqmax}
\begin{ruledtabular}
\begin{tabular}{l l l l l }
\hline
&\multicolumn{2}{c}{$\XtoS$} & \multicolumn{2}{c}{$\StoN$} \cr
$m_{ud}$ 
& \multicolumn{1}{c}{$q_{\rm max}^2$ [${\rm GeV}^2$]}
& $|f^{\rm ren}_S(q_{\rm max}^2)|$ 
& \multicolumn{1}{c}{$q_{\rm max}^2$ [${\rm GeV}^2$]}
& $|f^{\rm ren}_S(q_{\rm max}^2)|$\cr
\hline
0.005 &$-0.0103(16)$   & 0.9879(71)
          &$-0.0360(30)$   & 1.0166(112)\cr
0.01   & $-0.0063(15)$ &  0.9795(55) 
          & $-0.0223(28)$ &  1.0108(39) \cr
0.02   & $-0.0019(4)$ & 0.9928(16) 
	  & $-0.0080(7)$ & 1.0013(6) \cr
\hline
\end{tabular}
\end{ruledtabular}
\end{table}

\begin{table}[ht]
\caption{
Results for $R_{\Delta f}$ in $({\rm GeV})^{-4}$.
}
\label{Tab:R_Df}
\begin{ruledtabular}
\begin{tabular}{l l l}
\hline
$m_{ud}$ 
&\multicolumn{1}{c}{$\XtoS$}  &\multicolumn{1}{c}{$\StoN$}  \cr
\hline
0.005 &$-0.386(159)$ & $-0.689(281)$ \cr
0.01   & $-0.706(160)$ & $-0.503(229)$ \cr
0.02   & $-0.501(114)$ & $-0.472(193)$ \cr
\hline
physical point (linear) & $-0.498(243)$ & $-0.717(398)$\cr
physical point (average) & $-0.524(77)$ & $-0.529(125)$ \cr
\hline
\end{tabular}
\end{ruledtabular}
\end{table}

The excellent agreement observed here between two different fitting procedures
indicates that the systematic uncertainty stemming from the small deviation of the strange quark mass
appears to be relatively small in the AGT fit, where we directly insert the physical kaon and pion masses
into Eq.~(\ref{Eq:AGTform}) with the weighted average of $R_{\Delta f}$ in order to determine $\tilde{f}_1(0)$
at the physical point. However, we conservatively quote the global fit results as our final estimates.
The differences between two determinations may be regarded 
as the reliability of the extrapolation to the physical point in our current uncertainty.
Hence our final results are 
\be
f_1(0)= \left\{
\begin{array}{ll}
+0.9732(66)(7)(5) & {\rm for}\;\XtoS\cr
-0.9698(106)(15)(36)& {\rm for}\;\StoN,
\end{array}
\right.
\ee
where the first error is statistical, and 
the second and third are estimates of the systematic errors due to our choice 
of $q^2$ interpolation and the reliability of the extrapolation to the physical point, respectively. 
Note that since we simulate at a single lattice spacing, the systematic error introduced by the lattice discretization
is not estimated there.

It is worth emphasizing that the signs of the second-order corrections on $f_1(0)$
are consistent with what was reported in earlier quenched lattice studies~\cite{{Guadagnoli:2006gj},{Sasaki:2008ha}}
and preliminary results from mixed action calculation~\cite{Lin:2008rb} and
$n_f=2+1$ dynamical improved Wilson fermion calculations~\cite{Gockeler:2011se}.
However, we recall that the tendency of the SU(3)-breaking correction observed here
disagrees with predictions of both the latest baryon ChPT result~\cite{Geng:2009ik}
and large $N_c$ analysis~\cite{{Flores-Mendieta:1998ii},{FloresMendieta:2004sk}}.

We additionally remark that the latter has received some criticism from Mateu and Pich~\cite{Mateu:2005wi}.
They pointed out that the large $N_c$ fit including second-order SU(3)-breaking effects on $f_1(0)$ becomes
unreliable within the present experimental uncertainties~\footnote{
Indeed, they use the common value of $\tilde{f}_1=0.99(2)$, which is regarded as an educated guess, 
for all five decay modes including the neutron beta decay in order to extract $|V_{us}|$ from hyperon beta decays 
within the large $N_c$ framework. 
Due to large uncertainty of $\tilde{f}_1$, the resulting value of $|V_{us}|=0.226(5)$ receives much larger error than
other determinations of $|V_{us}|$.}.

\begin{table*}[!t]
\caption{
Results for $[f_1(0)/f^{\rm SU(3)}_1(0)]_{\XtoS}$
and $[f_1(0)/f^{\rm SU(3)}_1(0)]_{\StoN}$, where $f^{\rm SU(3)}_1(0)=+1$ for $\XtoS$ and 
$f^{\rm SU(3)}_1(0)=-1$ for $\StoN$. 
The final column gives the global fit results for the physical strange mass. 
}
\label{Tab:FinalResults}
\begin{ruledtabular}
\begin{tabular}{cclll ll}
\hline
&&  \multicolumn{3}{@{}c@{}}{$m_{ud}$}
& \multicolumn{2}{@{}c@{}}{physical point} \cr
&$q^2$ interpolation& 0.005 & 0.01 & 0.02 & AGT fit & Global fit\cr
\hline
$[f_1(0)/f^{\rm SU(3)}_1(0)]_{\rm \XtoS}$ & monopole & 0.9808(79) & 0.9742(58)& 0.9914(19) & 0.9737(39) &0.9732(66)\cr
& quadratic & 0.9806(81) & 0.9744(57) & 0.9910(19) & 0.9730(38) & 0.9731(67) \cr
\hline
$[f_1(0)/f^{\rm SU(3)}_1(0)]_{\rm \StoN}$ & monopole & 0.9656(140) & 0.9816(84) & 0.9919(33) & 0.9734(63) & 0.9698(106)\cr
& quadratic & 0.9641(140) & 0.9870(74) & 0.9918(31) & 0.9759(57) & 0.9748(99) \cr
\hline
\end{tabular}
\end{ruledtabular}
\end{table*}

\section{Summary}

We have studied the flavor SU(3)-breaking effect on hyperon vector coupling $f_1(0)$
for the $\Xi^0\rightarrow \Sigma^+$ and $\Sigma^{-}\rightarrow n$ decays in (2+1)-flavor QCD using 
domain wall quarks. We have observed that the second-order correction on $f_1(0)$ is still {\it negative} 
for both decays at simulated pion masses of $M_\pi=330-558$ MeV. 
The size of the second-order corrections observed here is also comparable to 
what was observed in our DWF calculations of $K_{l3}$ decays~\cite{Boyle:2007qe}.
Using the best estimate of $|V_{us}|=0.2254(6)$ with imposing CKM unitarity~\cite{Antonelli:2010yf},
we then predict the values $|V_{us}f_1(0)|_{\XtoS}=0.2194(8)_{V_{us}}(15)_{f_1}$ and $|V_{us}f_1(0)|_{\StoN}=0.2186(8)_{V_{us}}(24)_{f_1}$.
The former is barely consistent with a single experimental result of $|V_{us}f_1(0)|_{\XtoS}=0.209(27)$~\cite{AlaviHarati:2001xk}, albeit with its large experimental error. However, the latter is slightly deviated from the currently available experimental result of $|V_{us}f_1(0)|_{\StoN}=0.2282(49)$~\cite{Cabibbo:2003cu} due to reaching
the value of $f_1(0)$ with an accuracy of less than one percent.
We plan to extend our research to evaluate the systematic uncertainty 
due to the lattice discretization error and also to decrease the reliance on the chiral extrapolation
using RBC/UKQCD 2+1 flavor DWF dynamical ensembles at a second, finer, lattice spacing with
simulated pion masses closer to the physical point.


\begin{acknowledgments}
It is a pleasure to acknowledge the technical help of C. Jung for numerical calculations on the IBM BlueGene/L supercomputer.
I would also like to thank W. Marciano and V. Mateu for useful comments. 
This work is supported by the JSPS Grants-in-Aid for Scientific Research (C)
 (No.~19540265), Scientific Research on Innovative Areas
 (No.~23105704) and the Large Scale Simulation Program No.09/10-02 (FY2010) of High Energy Accelerator Research Organization (KEK). 
Numerical calculations reported here were carried out at KEK supercomputer system
and also on the T2K supercomputer at ITC, University of Tokyo.
\end{acknowledgments}



\begin{thebibliography}{99}



\bibitem{Sachrajda:2011tg} 
  C.~Sachrajda,
  PoS LATTICE {\bf 2010}, 018 (2010).
  
\bibitem{BM}
  E.~Blucher and W.~J.~Marciano,
  ``$V_{ud}$, $V_{us}$, Cabibbo Angle, and CKM Unitarity,''
  in J.~Beringer {\it et al.}  (Particle Data Group),
  Phys.\ Rev.\ D {\bf 86}, 010001 (2012).
    
\bibitem{Cabibbo:2003cu}
  For a review of hyperon beta decays, see 
  N.~Cabibbo, E.~C.~Swallow and R.~Winston,
  \emph{Ann.\ Rev.\ Nucl.\ Part.\ Sci.}  {\bf 53}, 39 (2003) and references therein.

\bibitem{Mateu:2005wi} 
  V.~Mateu and A.~Pich,
  JHEP {\bf 0510}, 041 (2005).
  
\bibitem{Garcia:1985xz} 
  A.~Garcia and P.~Kielanowski,
  Lect.\ Notes Phys.\  {\bf 222}, 1 (1985).

\bibitem{Gaillard:1984ny}
  J.~M.~Gaillard and G.~Sauvage,
  \emph{Ann.\ Rev.\ Nucl.\ Part.\ Sci.}  {\bf 34}, 351 (1984).


\bibitem{Guadagnoli:2006gj}
  D.~Guadagnoli, V.~Lubicz, M.~Papinutto and S.~Simula,
  \emph{Nucl.\ Phys.\  B} {\bf 761}, 63 (2007).
  
\bibitem{Sasaki:2008ha}
  S.~Sasaki and T.~Yamazaki,
  \emph{Phys.\ Rev.\  D} {\bf 79}, 074508 (2009).
    
\bibitem{Weinberg:1958ut}
  S.~Weinberg,
  \emph{Phys.\ Rev.\ } {\bf 112}, 1375 (1958).
  
\bibitem{Ademollo:1964sr}
  M.~Ademollo and R.~Gatto,
  \emph{Phys.\ Rev.\ Lett.}  {\bf 13}, 264 (1964).
  
\bibitem{Donoghue:1981uk}
  J.~F.~Donoghue and B.~R.~Holstein,
  \emph{Phys.\ Rev.\  D}{\bf 25}, 206 (1982).

\bibitem{Donoghue:1986th}
  J.~F.~Donoghue, B.~R.~Holstein and S.~W.~Klimt,
  \emph{Phys.\ Rev.\ D} {\bf 35}, 934 (1987).

          
\bibitem{Schlumpf:1994fb}
  F.~Schlumpf,
  \emph{Phys.\ Rev.\ D} {\bf 51}, 2262 (1995).

\bibitem{Flores-Mendieta:1998ii}
  R.~Flores-Mendieta, E.~Jenkins and A.~V.~Manohar,
  \emph{Phys.\ Rev.\ D} {\bf 58}, 094028 (1998).

\bibitem{Villadoro:2006nj}
  G.~Villadoro,
  \emph{Phys.\ Rev.\  D} {\bf 74}, 014018 (2006).

\bibitem{Lacour:2007wm}
  A.~Lacour, B.~Kubis and U.~G.~Meissner,
  \emph{JHEP} {\bf 0710}, 083 (2007).
  
\bibitem{Geng:2009ik}
  L.~S.~Geng, J.~Martin Camalich and M.~J.~Vicente Vacas,
  \emph{Phys.\ Rev.\  D} {\bf 79}, 094022 (2009).



\bibitem{Allton:2008pn}
  C.~Allton {\it et al.}  [RBC-UKQCD Collaboration],
  \emph{Phys.\ Rev.\  D} {\bf 78}, 114509 (2008).
  
  
\bibitem{Yamazaki:2008py}
  T.~Yamazaki {\it et al.}  [RBC+UKQCD Collaboration],
  \emph{Phys.\ Rev.\ Lett.}  {\bf 100}, 171602 (2008).

\bibitem{Yamazaki:2009zq}
  T.~Yamazaki {\it et al.} [RBC+UKQCD Collaboration],
  \emph{Phys.\ Rev.\  D} {\bf 79}, 114505 (2009).
  
\bibitem{Aoki:2010xg}
  Y.~Aoki {\it et al.} [RBC+UKQCD Collaboration],
  \emph{Phys.\ Rev.\  D} {\bf 82}, 014501 (2010).

\bibitem{Sasaki:2011hu}
  S.~Sasaki,
  AIP\ Conf.\ Proc.\ {\bf 1388} (2011) 443-446; 
  arXiv:1102.4934 [hep-lat].


\bibitem{Sasaki:2003jh} 
  S.~Sasaki, K.~Orginos, S.~Ohta and T.~Blum, [RIKEN-BNL-Columbia-KEK Collaboration],
  Phys.\ Rev.\ D {\bf 68}, 054509 (2003).
  
\bibitem{Lin:2008rb}
  H.~W.~Lin,
  Nucl.\ Phys.\ Proc.\ Suppl.\  {\bf 187}, 200 (2009).

\bibitem{Gockeler:2011se}
  M.~Gockeler {\it et al.}  [QCDSF Collaboration and UKQCD Collaboration],
  PoS {\bf LATTICE2010}, 165 (2010).
    
\bibitem{FloresMendieta:2004sk} 
  R.~Flores-Mendieta,
  Phys.\ Rev.\ D {\bf 70}, 114036 (2004).

\bibitem{Boyle:2007qe}
  P.~A.~Boyle {\it et al.} [RBC+UKQCD Collaboration],
  \emph{Phys.\ Rev.\ Lett.}  {\bf 100}, 141601 (2008).
  
  

  
\bibitem{Antonelli:2010yf} 
  M.~Antonelli, V.~Cirigliano, G.~Isidori, F.~Mescia, M.~Moulson, H.~Neufeld, E.~Passemar, M.~Palutan, {\it et al.},
  Eur.\ Phys.\ J.\ C {\bf 69}, 399 (2010).

\bibitem{AlaviHarati:2001xk} 
  A.~Alavi-Harati {\it et al.}  [KTeV Collaboration],
  Phys.\ Rev.\ Lett.\  {\bf 87}, 132001 (2001).
 \end{thebibliography}
\end{document}